\documentclass[journal]{IEEEtran}
\IEEEoverridecommandlockouts
\usepackage{amsmath,epsfig,color,url}
\usepackage{amsfonts}
\usepackage{graphicx}
\usepackage{fancyhdr}
\usepackage{array}

\newcommand{\poww}{PowWow}
\newcommand{\flopoco}{FloPoCo}
\newcommand{\igloo}{IGLOO}

\IEEEpubid{\makebox[\columnwidth]{978-1-4673-2921-7/12/\$31.00 \copyright2012 IEEE} \hspace{\columnsep}\makebox[\columnwidth]{ }}

\graphicspath{{./fig/}}

\begin{document}

\title{Hardware Implementation of the GPS authentication}

\author{\IEEEauthorblockN{Micka\"el Dardaillon, C\'edric Lauradoux and Tanguy Risset}\\
\IEEEauthorblockA{Universit\'e de Lyon, INRIA,\\
INSA-Lyon, CITI-INRIA, F-69621, Villeurbanne, France \\
Emails: \{mickael.dardaillon, tanguy.risset\}@insa-lyon.fr, cedric.lauradoux@inria.fr
\thanks{This work is partially supported by R\'egion Rh\^one Alpes ADR 11 01302401.}}
}

\maketitle

\begin{abstract}
In this paper, we explore new area/throughput trade-offs for the 
Girault, Poupard and Stern authentication protocol (GPS). This 
authentication protocol was selected in the NESSIE competition 
and is even part of the standard ISO/IEC~9798. The originality 
of our work comes from the fact that we exploit a fixed key to 
increase the throughput. It leads us to implement GPS using the 
Chapman constant multiplier. This parallel implementation is 40 times 
faster but 10 times bigger than the reference serial one. We propose 
to serialize this multiplier to reduce its area at the cost of lower 
throughput. Our hybrid Chapman's multiplier is 8 times faster but 
only twice bigger than the reference. Results presented here allow 
designers to adapt the performance of GPS authentication to their 
hardware resources.  The complete GPS prover side is also integrated 
in the network stack of the {\poww} sensor which contains an Actel 
{\igloo} AGL250 FPGA as a proof of concept. 
  
\end{abstract}


\section{Introduction}

A current trend in cryptography is the design of primitives with a low-cost implementation. 
\textit{Lightweight cryptography}~\cite{Eisenbarth2007} is interested in metrics such as the circuit area, energy consumption or code size. The Girault, Poupard and Stern authentication protocol (GPS) is 
one of the oldest lightweight primitives which has been recommended by the European project 
NESSIE~\cite{Baudron2001} and appeared in the ISO/IEC~9798-5 standard~\cite{iso}. GPS is a 
reference in the domain of lightweight authentication and the existing implementations aim to 
reduce as much as possible the area. In this paper, we explore the trade-off between 
speed and area in the implementation of GPS. 

Implementing GPS consists  in designing an adder and a multiplier. The latter is the most critical part in a GPS core. Three typical strategies 
can be followed to implement this core: \textit{serial}, \textit{parallel} and a trade-off 
between the previous two that we designate by \textit{hybrid}. The serial implementation based on
\textit{shift-and-add} has been already covered in~\cite{Mcloone2007} and is the baseline for area. 
The other two strategies have not yet been studied. The parallel implementation is the traditional 
method to achieve the best speed at the drawback of a large area. The hybrid implementation in our paper is in fact a serialization of the parallel approach that takes advantage of hardware repetition. It provides a trade off between serial/parallel approaches by executing chunks of parallel multiplication.

To implement a parallel implementation of GPS, we face a big challenge: implementing parallel multiplier with large variable operands is out-of-question. To overcome this limitation, we have specialized our parallel and hybrid implementations for a specific key. The complexity of our multiplier is therefore reduced to the implementation of a \textit{multiplier by a constant}~\cite{Wirthlin2004}.

We attempt to integrate our different implementations in the network stack of the wireless sensor platform
{\poww}. This platform embeds an Actel {\igloo} FPGA with a 8 MHz clock cycle. Our simulation results for a security level of 128-bit show that the parallel implementation of GPS with a  secret takes 1$\mu s$ against 42$\mu s$ for the existing serial implementation at 8~MHz. However, it does not fit within the FPGA. The integration of GPS into our platform is only possible for hybrid and serial implementation. 

We note that the basic GPS protocol requires a 320$\mu s$ response time, but McLoone and Robshaw circumvent this difficulty by using a different protocol in their article~\cite{Mcloone2007} with a 18$m s$ latency. Our implementation fulfil the 320$\mu s$ constraint for all architectures at 8~MHz, and would help reach this goal at a lower frequency for parallel and hybrid implementations.

The rest of this paper is organized as follows. The GPS protocol is reminded in Section~\ref{sec:gps}. The Section~\ref{sec:serial} describes the existing serial implementation of GPS. The main contributions of the paper are given in Section~\ref{sec:parallel}  and~\ref{sec:hybrid} with the description of the parallel and hybrid implementation. The performances are compared and analyzed in Section~\ref{sec:comp}. 
 
\section{GPS protocol}\label{sec:gps}

The GPS authentication protocol~\cite{Baudron2001} is an interactive zero-knowledge authentication protocol initially proposed by Girault, Poupard, and Stern~\cite{Girault2006}. It provides provable security based on the composite discrete logarithm problem. It also combines short transmissions and minimal on-line computation, using precomputed ``coupons''.  This protocol has been selected in the NESSIE portfolio~\cite{nessie} and it is mentioned in the ISO/IEC~9798-5 Clause~8~\cite{iso} as a reference. Throughout the paper, we implicitly refer GPS as this variant ``with coupons''.

\paragraph{Description} The parameters used in this protocol are the following:
\begin{itemize}
\item $S$, $C$, $D$ are public integers, where $|S| \approx 180$ bits\footnote{In all the paper we use the notation $|X|$ for {\em the size in bits of number $X$}}, $|C| = 32$ and $|D| = |S| + |C| + 80$,
\item $n = p \times q$ is a public composite modulus, where $p$ and $q$ are secret primes, $|n| = 1024$, $|p| = |q| = 512$,
\item $g$ is an element of $\mathbb{Z}_n^*$,
\item $\Phi = (C-1) \times (S-1)$,
\item $s \in [0,S[$ and $I = g^{-s} \mod n$,
\item a coupon $i$ is a couple $(r_i, x_i = g^{r_i} \mod n)$, where $r_i \in [0, D[$ is a random number.
\end{itemize}

At the beginning, the prover $P$ has a unique identifier $Id_P$, a unique pair of keys (the private $s$ and the public $I$) and a set of coupons $c$ computed by a higher trusted entity. The verifier $V$ knows the prover's identifier and public key. 

\begin{figure}[htb]
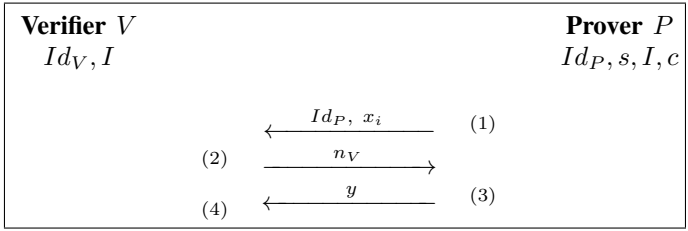

	\centering
	\framebox[0.5\textwidth]{
	\begin{tabular}{p{4cm}cp{4cm}}
	\multicolumn{1}{c}{\textbf{Verifier $V$}} & & \multicolumn{1}{c}{\textbf{Prover $P$}}\\
	\multicolumn{1}{c}{$Id_V, I$} & & \multicolumn{1}{c}{$Id_P, s, I, c$}\\ \\
	& $\xleftarrow{\ \ \ \ Id_P,\ x_i\ \ \ \ \ }$ & $^{(1)}$\\
	\multicolumn{1}{r}{$^{(2)}$} & $\xrightarrow{\ \ \ \ \ \ \ n_V\ \ \ \ \ \ \ }$ & \\
	\multicolumn{1}{r}{$_{(4)}$}& $\xleftarrow{\ \ \ \ \ \ \ \ y\ \ \ \ \ \ \ \ }$ & $^{(3)}$\\
	\end{tabular}}
	\caption{The GPS protocol.}\label{fig:gps}
\end{figure}

GPS, depicted in Fig.~\ref{fig:gps}, works as follows.
\begin{enumerate}
\item[(1)] The prover $P$ chooses a coupon $(r_i, x_i)$, and sends its identifier $Id_P$ and $x_i$ to the verifier $R$.
\item[(2)] The verifier answers a challenge $n_V$ randomly chosen in the interval $[0, C[$.
\item[(3)] The prover computes $y = r_i + n_V \times s$, and sends $y$ to the verifier.
\item[(4)] The verifier checks if:
\begin{itemize}
\item $g^y \times I^{n_V} \mod n = x_i$,
\item $y \in [0, D + \Phi[$.
\end{itemize}
\end{enumerate}

The arguments supporting the security of GPS can be found in~\cite{Baudron2001,Girault2006}. They are not 
included because the security of GPS is not affected by our results. 

In the remaining of the paper, $s$ is named \textit{the secret}, $n_V$ is named \textit{the challenge} and $r_i$ \textit{the commitment} (see~\cite{Baudron2001}).

\paragraph{Existing implementation} GPS authentication has been designed for 
constraint embedded systems such as smart cards, RFID or sensors networks. The critical part 
for the implementation is on the prover side assuming that the verifier (RFID reader or a base station) suffers from less restriction. 

Two steps are critical for the prover: the computation of $x_i$ (Step~(1)) and $y$ 
(Step~(3)). The computation of $x_i$ is the most expensive one (exponentiation) 
but the prover has nothing to do thanks to the coupons (pre-computation). Therefore, 
the last remaining cost is the Step~(3) which consists of the multiplication by $s$ 
and the addition of $r_i$. The multiplication by $s$ is the most complex due to 
the size of the multiplicands. 

GPS was first designed for smart cards~\cite{Girault2006}. McLoone and Robshaw 
proposed in~\cite{Mcloone2007,McLoone2007b} the first hardware implementation of GPS. Their 
solution is based on the shift-and-add algorithm for multiplication. It 
offers a very small hardware footprint. This implementation is described
in Section~\ref{sec:serial} and it serves as the reference in our comparison.  

Girault and Lefranc~\cite{Girault2004} proposed a variant  of GPS which exploits 
low Hamming weight secret $s$. The multiplication is transformed in an addition 
when the secret is chosen properly. This variant can significantly reduce the 
cost of GPS. However, this solution was subsequently attacked 
in~\cite{Coron2005}, and it is not considered in this work.

In the following sections, the different architectures for GPS are
explored. Two criteria are examined: \textit{area} and
\textit{speed}. A cryptographic design can be tuned for a specific key
or support any value for the key. Throughout this paper, fixed-key and
variable-key implementation are considered.

\section{Serial implementation}\label{sec:serial}

\begin{figure}[h]
\centering
$\begin{array}{l c l r}
&&& 101001 \\
&& \times & 110 \\ \cline{3-4}
1 \times 101001 \times 100 & \rightarrow & + & 101001\hspace{0.08cm}.\hspace{0.08cm}. \\
1 \times 101001 \times 10 & \rightarrow & + & 101001\hspace{0.08cm}. \\
0 \times 101001 & \rightarrow & & 000000 \\  \cline{3-4}
&&&   11110110
\end{array}$
\caption{Shift-and-add classical binary multiplication.}\label{fig:school_bin_product}
\end{figure} 

McLoone and Robshaw have proposed in~\cite{Mcloone2007} the reference serial 
implementation of GPS. This architecture is based on the classical shift-and-add multiplication~(Fig.~\ref{fig:school_bin_product}). The core of the design is a 16-bit adder. The product $n_V\times s$ is decomposed into a succession of 16-bit additions. The same 16-bit adder is re-used to perform the final addition $ n_V\times s+r_i$ in 16-bit chunks.

\begin{figure}[!ht]
	\centering
	\includegraphics[width=.5\textwidth]{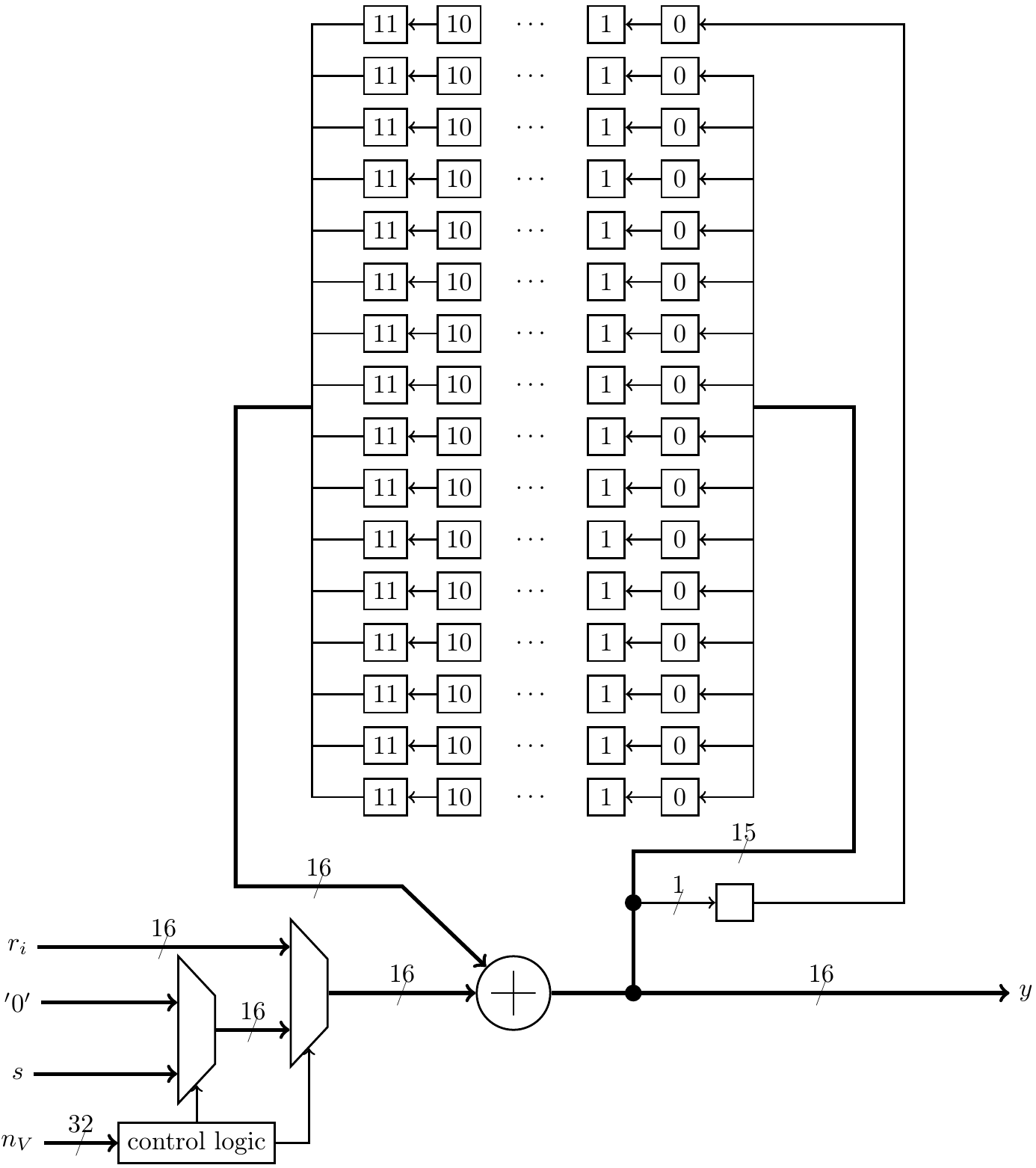}
	\caption{GPS serial implementation for a 128-bit secret and a 16-bit  adder (from \cite{Mcloone2007}).}
\label{fig:GPS_serial}
\end{figure}

The architecture of McLoone and Robshaw is presented on Fig.~\ref{fig:GPS_serial}. The control logic block process the challenge $n_V$ bitwise and drives the multiplexers. The first multiplexer selects 16 bits of the secret $s$ or $0$, depending on the challenge bit. The adder sums the multiplexer data with the previous results stored in the shift register. The final addition of $r_i$ with the result is controlled by the second multiplexer. The same hardware is re-used for this addition. The data is processed sequentially by 16-bit blocks to obtain a serial architecture. The balance between occupied area and execution time is set by the data bus size. Doubling the data bus size halve the required run cycles, while increasing the adder and the multiplexer size.  

We have implemented the serial architecture of  McLoone and Robshaw independently, and we get area results close to their original article for the different security parameters (see Table~\ref{tab:GPSvsarticle}). However, McLoone and Robshaw results only take into account the computing hardware. We give results for a complete implementation of the serial multiplier, including memories for $n_V$, $r_i$, $s$ and $y$.

\section{Parallel implementation}\label{sec:parallel}

A multiplier can be implemented using lookup tables (ROM). Let us consider two variable operands $n_V$ and $s$, of size $|n_V|$ and $|s|$. The product $n_V\times s$ has a $|n_V|+|s|$ bits width. A lookup table approach requires to store: $$2^{|n_V|+|s|}\times (|n_V|+|s|) \text{ bits.}$$ While being attractive for small values of $n_V$ and $s$, lookup tables are clearly not practicable for GPS, \textit{i.e.} $|n_V|\geq 32$.

A first attempt to reduce this cost consists in fixing $s$. This hypothesis 
implies that our implementation is dedicated to a given key. In most cryptographic 
implementations, the key can be updated. To our knowledge, there are two exceptions. 
FPGA supplier provides encryption scheme with fixed key to protect the users bit-stream. White-box cryptography~\cite{Wyseur2011} also uses implementation with fixed key but to obfuscate 
the key into the code. Our goal is to show that implementation with fixed key can provide benefits in term of throughput.

Using this assumption, the  cost of a table lookup multiplier can be reduced to: $$2^{|n_V|}\times
(|n_V|+|s|) \text{ bits.}$$ This approach is still not reasonable for
GPS.

To further reduce the memory size, Chapman proposed
in~\cite{Chapman1996} the KCM method (for {\em k constant
  multiplier}) to decompose the computation into partial products. Let
us explain Chapman's KCM method in decimal basis 
(Fig.~\ref{fig:constant_product}) for the multiplication of $953 \times
x$.  If one can store in look-up tables (LUT) the nines values
$1\times 953$, $2\times 953$,\ldots, $9\times 953$, then one can
compute the whole product by simply adding stored values as indicated in the  Fig.~\ref{fig:constant_product} for $953 \times 482$.

\begin{figure}[h]
\centering
$\begin{array}{l c l r}
&&& 953 \\
&& \times & 482 \\ \cline{3-4}
4 \times 953 \times 100 & \rightarrow & + & 3812\hspace{0.08cm}.\hspace{0.08cm}. \\
8 \times 953 \times 10 & \rightarrow & + & 7624\hspace{0.08cm}. \\
2 \times 953 & \rightarrow & & 1906 \\ \cline{3-4}
&&&   459346
\end{array}$
\caption{Chapman KCM principle for multiplication by 953 in decimal basis.}\label{fig:constant_product}
\end{figure}

For an hardware design, KCM uses $2^{\ell}$ basis rather than the
decimal basis. The choice of $\ell$ depends on the trade-off between
the cost (memory and adder) and the technology characteristics
(number of inputs per lookup table). For example, a classic Xilinx Virtex 4 as 4-input LUT, and our Actel Igloo as 3-input LUT. First, $2^\ell$ values need to be
stored in each table, hence for a  total memory size of: 
$$\frac{|n_V|}{\ell} \times 2^{\ell}\times (|s|+\ell) \text{
  bits.}$$ The results obtained from these tables are combined by
$\frac{|n_V|}{\ell}-1$ adders of $|s|+\ell$ bits.

\begin{figure}[!ht]
\centering
	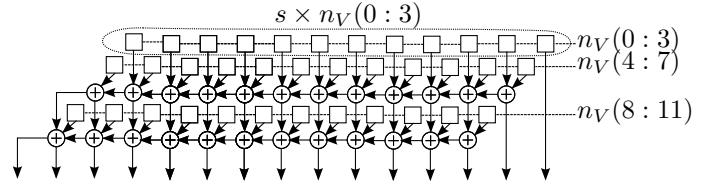
	\caption{KCM constant multiplier (adapted from~\cite{Wirthlin2004}).}\label{fig:KCM}
\end{figure}

This architecture is illustrated on Fig.~\ref{fig:KCM}
for $|s|=8$, $|n_V|=12$ and $\ell=4$. Each lookup table takes one part
of $n_V$ as an input. The partials results are combined using 2 adders
to produce the final result. Note that the left shifts are done by
positioning the partial results and have no gate cost. Note also that
the result is obtained in a single cycle, with a very long critical
path that could slow down the clock cycle of the circuit. To cope with this 
clock cycle problem, the designer can pipeline the circuit, increasing 
latency but maintaining a throughput of 1 result per clock cycle.

For our GPS implementation of the parallel approach, we used a  KCM
operator generated by the {\flopoco}
library\footnote{\url{http://flopoco.gforge.inria.fr}}. This library
is an open source project for non-standards arithmetic operators. It provides in
particular several integer constant
multipliers~\cite{Brisebarre2008}. All operators are pipelined to run
at high frequency (default setting at 300 MHz on Virtex FPGA).

\begin{figure}[!ht]
\centering
	\includegraphics[width=.4\textwidth]{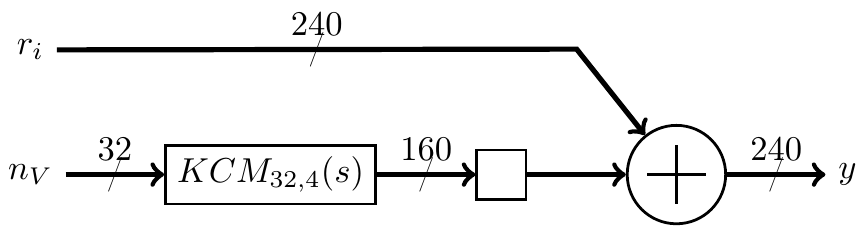}
	\caption{Parallel implementation of GPS (128 bits).}\label{fig:GPS_parallel}
\end{figure}

The Fig.~\ref{fig:GPS_parallel} illustrates the parallel
implementation of GPS used in our design. The notation $KCM_{32,4}(s)$
stands for {\em Chapman constant multiplier of a 32-bit number with the
constant $s$, using 4-input LUT}.  The product is computed by the constant multiplier core
generated with {\flopoco}, with the secret $s$ as a constant. The
result $s \times n_V$ is added to $r_i$. 

This architecture is the fastest in terms of speed using the property
that the secret $s$ is constant to reduce the size of the
implementation. The performance results of the parallel approach are
presented further in the paper, but the design could not fit on a
small embedded FPGA such as the {\igloo} AGL250 used in {\poww}, hence the need of
a trade-off between the serial and parallel approach.

\section{Hybrid implementation}\label{sec:hybrid}

\begin{figure}[h]
\centering
	\includegraphics[width=.5\textwidth]{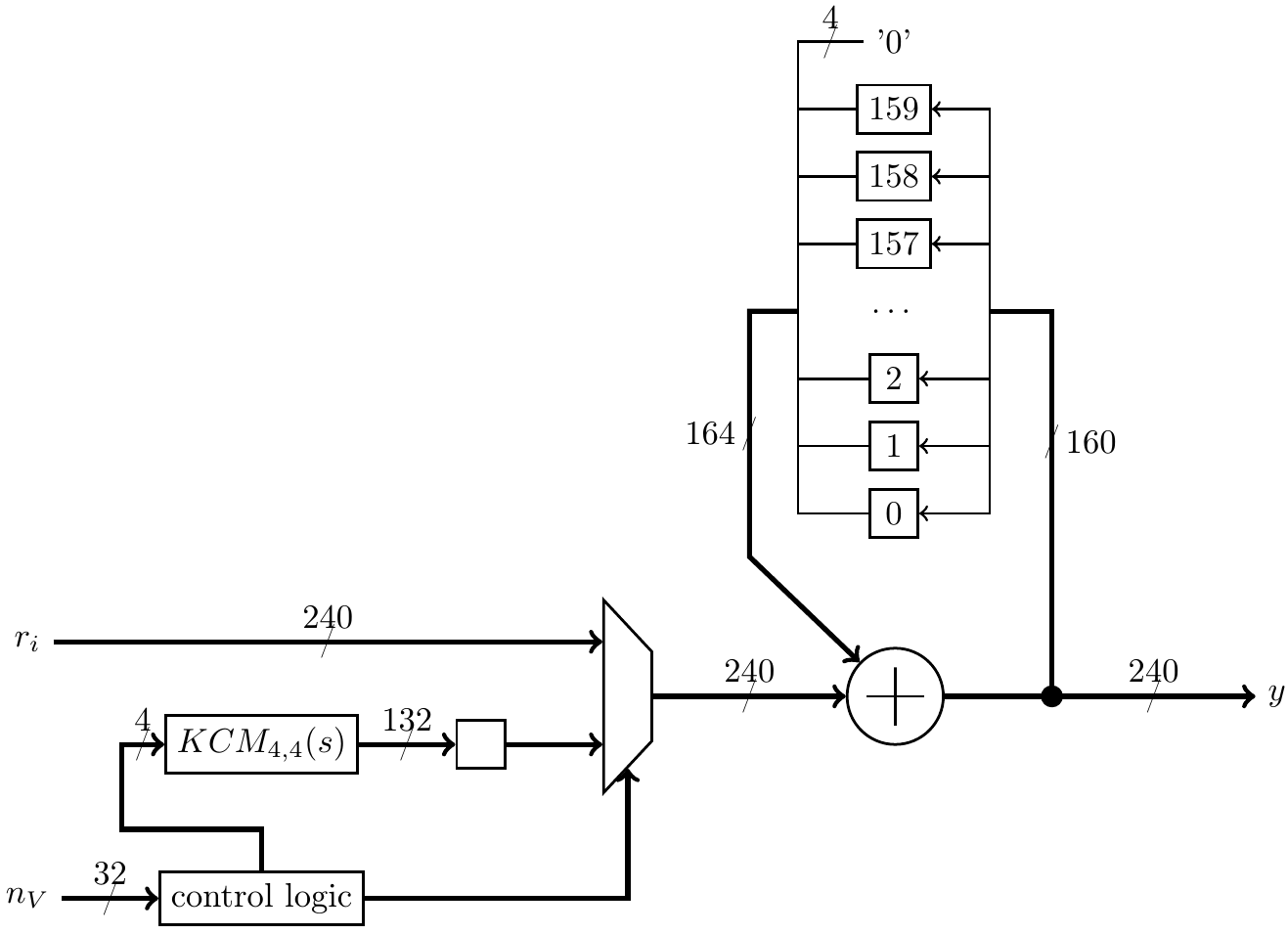}
	\caption{Hybrid KCM architecture (128 bits).}\label{fig:hybrid_kcm}
\end{figure}

The idea of the hybrid implementation is to serialize the KCM architecture of Fig.~\ref{fig:KCM} {\em horizontally}. Indeed, each horizontal row of LUT is the same: it contains all the values of $s$ multiplied by $0,1,\ldots, 2^{\ell}-1$. Hence if we sequentialize the addition of one row with the others we may use a single LUT of size $2^{\ell}\times|s| + \ell$. The challenge $n_V$ is sent by blocks of four bits to drive the LUT, the output of the LUT is added to the accumulation register and then shifted by  four bits for the next addition.

The architecture, illustrated on Fig.~\ref{fig:hybrid_kcm}, uses one KCM table of $2^{\ell}\times (|s|+\ell) \text{ bits}$ to compute the product between $\ell$ bits of the challenge and the constant $s$. The result is accumulated and shifted by $\ell$ bits each cycle. The adder is reused to sum the result of the multiplication with $r_i$. All the computation is done using a parallel adder. To further reduce the critical path, the adder could also be sequentialized with a sequence of smaller adders as it was done for the serial approach in section~\ref{sec:serial}.

\section{Experimental platform}\label{sec:exp}
We performed our implementation on the wireless sensor platform
{\poww}\footnote{\url{http://powwow.gforge.inria.fr}}. This platform is
specifically designed to be energy efficient~\cite{Berder2010}, and
offers a constraint hardware platform in terms of area and
time. It uses the Actel {\igloo} AGL250 FPGA for control and a Texas
Instrument CC2420 chip for RF communications. Actel {\igloo} uses
non volatile flash technology for FPGA dedicated to low power.

We integrated the GPS authentication into the network stack of the
platform. The physical layer is based on the 802.15.4 standard with a CSMA/CA access provided
by the CC2420 and controlled by the FPGA. It supports a simple point to point
access with the verifier in order to process the whole GPS authentication (see Fig.~\ref{fig:gps} for details).
In our set-up, the verifier is a computer interfaced with a Senslab\footnote{\url{http://www.senslab.info}}
node for telecommunications. 

\begin{figure}[ht]
\centering
	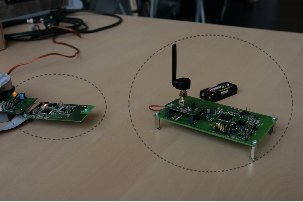
	\caption{{\poww} running a wireless authentication.}\label{fig:demonstrator}
\end{figure}

The implementation was verified by authenticating the prover to the verifier 8 times
in a row, and with different keys on separate synthesis, using vectors generated by 
GMP\footnote{\url{http://gmplib.org}} as reference. The serial architecture was evaluated
with several challenge and key sizes ($|s|=\{128,256,512\} , |n_V| = \{16,20,32\}$).
The hybrid architecture could be implemented on the platform for a key size up to 256 bits. 
The experimental set-up is illustrated on Fig.~\ref{fig:demonstrator}.

In order to validate our work, we have implemented independently the 
serial approach of GPS to check that our results were compatible with those 
of McLoone and Robshaw in~\cite{Mcloone2007}. The synthesis was done on the Cadence
ATL Compiler and normalized to a NAND size to be consistent with~\cite{Mcloone2007}. 
The Table~\ref{tab:GPSvsarticle} compares our area results with the existing ones. The 
difference is below 5~\% for a serial implementation with an 8-bit adder and below 10~\% 
with an 16-bit adder. We consider these differences as acceptable: 
our serial implementation can be used as a benchmark to compare the different architectures.

\begin{table}[!ht]
\centering
\caption{Area comparison between our serial implementation and the one of McLoone and Robshaw~\cite{Mcloone2007}.}\label{tab:GPSvsarticle}
\begin{tabular}{c c c c c}
Secret & Challenge & \multicolumn{2}{c}{Area (NAND)} & Difference \\
(bits) & (bits) & \cite{Mcloone2007} & our work & \\ \hline
8 bits adders \\
160 & 32 & 1541 & 1505 & 2.31\% \\
128	& 32 & 1327	& 1320 & 0,54\% \\
160	& 20 & 1486 & 1413 & 4,89\% \\
128	& 20 & 1286 & 1231 & 4,28\% \\
160	& 8  & 1371 & 1341 & 2,21\% \\
128	& 8  & 1167 & 1163 & 0,37\% \\ \hline
16 bits adders \\
160	& 32 & 1642 & 1594 & 2,90\% \\
128	& 32 & 1411 & 1403 & 0,54\% \\
160	& 20 & 1642 & 1502 & 8,53\% \\
128	& 20 & 1411 & 1314 & 6,90\% \\
160	& 8  & 1511 & 1395 & 7,65\% \\
128	& 8  & 1298 & 1205 & 7,16\%
\end{tabular}
\end{table}

\section{Implementation results}\label{sec:comp}

From Section~\ref{sec:serial} to~\ref{sec:hybrid}, three different approaches 
to implement GPS have been described to obtain a low area footprint, an high throughput or a trade-off 
between them. We now aim to verify these expectations, compare our implementations 
quantitatively and determine the influence of the secret size. 

The three approaches discussed above were written in VHDL and different synthesis were made. 
All the results are given for a synthesis and a mapping performed using Actel tools and targeting Actel {\igloo} AGL250. The targeted clock cycle was fixed for all impementations at 8 MHz, frequency used on the {\poww} platform. During our experiments, we have aligned systematically the bus size on the adder size. Moreover, 
we evaluated the overall footprint, including all the required memories. This is an important point, 
as it turns out that memories are occupying much more space than computing parts. It also permits to 
illustrate the gains obtained by using a constant multiplier, which includes the constant memory.

\begin{figure}[!ht]
\centering
    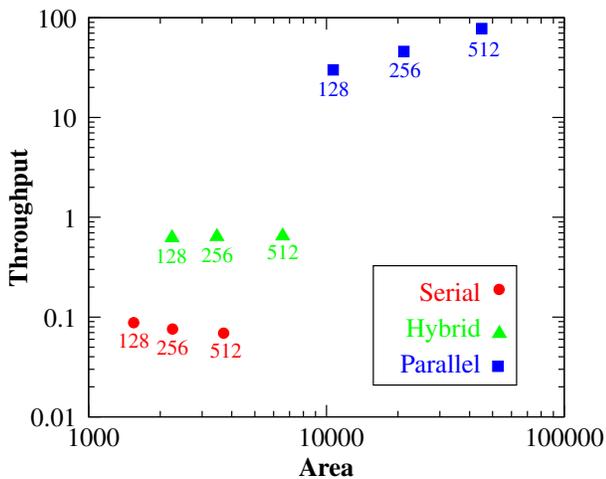	
	\caption{Area vs throughput for a 32-bit challenge (log scale) and different secret sizes 
	($|s|=\{128,256,512\}$).}\label{fig:plot}
\end{figure}

We plotted the area and throughput for the three implementations on Fig.~\ref{fig:plot}.
This illustrates the three different balances between area and throughput achieved by the three implementations.
Moreover, these balances are maintained for the different levels of security, and both the area and throughput can be represented by a linear approximation in function of the secret size ($y=a*|s|+b$).
We observe that the area is increasing with the security size for the serial ($a=5,6$), the parallel ($a=89,8$) and the hybrid ($a=11,3$ ) implementation, with the best scaling for the serial approach in regards of the area.
In terms of throughput, the serial implementation as a negative slope ($a=-46,3*10^{-6}$), contrary to the parallel  ($a=125*10^{-3}$) and hybrid ($a=61,9*10^{-6}$) implementations, which gets a better throughput by using a larger secret $|s|$.

\begin{table}[!ht]
	\centering
	\caption{Area, latency and throughput comparison between implementations for a 32-bit challenge and different secret sizes.}\label{tab:ratio}
	\begin{tabular}{c@{\hspace{.8cm}} c@{\hspace{.8cm}} c@{\hspace{.8cm}} c@{\hspace{.8cm}}}
	Secret & \multicolumn{3}{c}{Implementation}\\

	(bits) &  Serial & Parallel & Hybrid  \\ \hline
	
    & \multicolumn{3}{c}{Area (core cells)} \\ \cline{2-4}
	128 & 1546 & 10676 & 2243 \\
	256 & 2253 & 21171 & 3467 \\
	512 & 3698 & 44978 & 6553 \\ \hline

     &  \multicolumn{3}{c}{Latency (cycles)} \\ \cline{2-4}
	128 & 339 & 8 & 48 \\
	256 & 603 & 12 & 72 \\
	512 & 1131 & 20 & 120 \\ \hline
	
    &  \multicolumn{3}{c}{Throughput (cycles/byte)} \\ \cline{2-4}
	128 & 0,088 & 30 & 0,625 \\
	256 & 0,076 & 46 & 0,639 \\
	512 & 0,069 & 76 & 0,650 \\
	\end{tabular}
\end{table}

The results shown in Table~\ref{tab:ratio} compares the three implementations in
terms of hardware area, run cycles and throughput (in cycles per byte of result).
We use the serial implementation as a reference for our comparison.
The parallel implementation has an area 10 times larger than the serial implementation.
For that cost, its pipeline structure provides a new result each cycle and a 40 times smaller latency when using multi-authentication.
The hybrid implementation offers a middle ground with an area less than doubled,
and an 8 times smaller latency.

We also explored the impact of the adder size on the area for the serial implementation in the Table~\ref{tab:GPS_bus_size}. To reduce the area, it would seems logical to choose the smallest adder. However, it impacts also the bus size and how the memory is addressed. Therefore, choosing the smallest adder is not necessarily  the best solution because it can imply  
an addressing overhead. The gain for the adder can be offset by the memory cost. This effect is highly dependent on the underlying technology. In our case, the 16-bit adder has a lower area than the 8-bit one.

\begin{table}[!ht]
	\centering
	\caption{Impact of the adder on the serial implementation area for a 32-bit challenge.}\label{tab:GPS_bus_size}
	\begin{tabular}{c  c  c  c}
	Adder & \multicolumn{3}{c}{Secret (bits)} \\
	(bits) & 128 & 256 & 512 \\ \hline
    & \multicolumn{3}{c}{Area (core cells)} \\ \hline
	8  & 1542 & 2270 & 3745 \\
	16 & 1546 & 2253 & 3698 \\
	32 & 1934 & 2632 & 4034 \\
	\end{tabular}
\end{table}

In order to be complete on the implementation footprint, we should also add the coupon footprint. By using a PRNG as suggested in the McLoone and Robshaw article~\cite{Mcloone2007}, one can reduce the footprint to 1000 NAND equivalent for storing 20 coupons, with another 1000 NAND for the PRNG. This would require about 2300 core cells, and is compatible with the serial and hybrid implementation for 128-bit and 256-bit secret size.

\section{Conclusion}

Several architectures  for GPS are presented in this work  offering different 
balances between area and throughput. With respect to this goal, our contribution is three-fold. First, we have shown that using a fixed key can enable new implementations possibilities. Second, we have serialized KCM to reduce its area cost. To our knowledge, this is the first time that such a trade-off is proposed for KCM. Third, we have integrated our GPS cores into a real platform ({\poww}) to enable nodes authentication in a wireless sensor network. By integrating the memory cost, we have a better view of the overall area needed for GPS. Two of our cores are compatible with the platform constraints. 

We have shown the benefits of having a fixed key for the implementation of GPS. In return, an adversary may exploit these features to mount \textit{side-channel attacks} (SPA, DPA\dots). While not considered in this work, it will be interesting to analyze how resistant are our implementations to these attacks. 

We focused on GPS but our results impact the implementation of more cryptosystems. In term of hardware architecture, GPS is very similar to WIPR~\cite{Oren2009} and  BlueJay~\cite{Saarinen2012}. These two cryptosystems are based on an early paper of Shamir~\cite{Shamir1994} which introduced \textit{randomized multiplication cryptosystem}. The cores of these proposals are an adder and a multiplier as for GPS. Our solutions can be adapted to implement WIPR, BlueJay and Shamir's scheme. In a future work, we will compare the implementations of these different schemes.

%
%
%

\section*{Acknowledgments}

The authors wants to thank Florent de Dinechin from Aric INRIA team for his help on constant multiplication and Romain Fontaine from the CAIRN INRIA team for his help on the RF-communication of {\poww} platform.

\bibliographystyle{IEEEtran}
\bibliography{gps}

\end{document}